\documentclass{elsart}

\usepackage{amssymb}
\usepackage{epsf}

\begin{document}

\begin{frontmatter}


\title{Possible astrophysical clues of dark matter }

\author{Alexander Kusenko}

\address{Department of Physics and Astronomy, UCLA, Los Angeles, CA
90095-1547\\ RIKEN BNL Research Center, Brookhaven National Laboratory,
Upton, NY 11973 }

\begin{abstract}

The supernova physics may provide a clue of the cosmological dark matter.
In the absence of new physics, the supernova calculations do not explain
the observed velocities of pulsars.  However, if there exists a singlet
fermion with mass in the 1--20~keV range and a small mixing with
neutrinos, this particle could be emitted asymmetrically from a cooling
neutron star in the event of a supernova explosion.  The asymmetry could
explain the long-standing puzzle of pulsar velocities.  The same particle
could be the dark matter.  Observations of X-ray telescopes, as well as a
future detection of gravitational waves from a nearby supernova can confirm
or rule out this possibility.
\end{abstract}

\end{frontmatter} 

\section{Introduction}
Very few hints exist as to the nature of cosmological dark matter.  We know
that none of the Standard Model particles can be the dark matter, and we
also know that the dark matter particles should either be weakly
interacting or very heavy (or both).  Theoretical models have provided
plenty of candidates.  For example, the supersymmetric extensions of the
Standard Model predict the existence of a number of additional particles,
which include two dark matter candidates: the lightest supersymmetric
particle (LSP) and the SUSY Q-balls.  These are plausible candidates, which
were discussed in a number of talks at this conference.  In the absence of
observational hints, one naturally relies on theoretical models in planning
experimental searches for dark matter.

There is, however, another astrophysical puzzle, which may provide a clue
as to the nature of dark matter.  The origin of observed pulsar
velocities~\cite{astro,astro_1} remains a mystery, but it can be explained
if a weakly interacting particle is emitted form a cooling neutron star
isotropically during the first seconds of the supernova explosion.  There
is an intriguing possibility that the same weakly interacting particle
makes up the dark mater.

Pulsar velocities range from 100 to
1600~km/s~\cite{astro,astro_1}. Their distribution  leans toward
the high-velocity end, with about 15\% of all pulsars having speeds over
(1000~km/s)~\cite{astro_1}.

Pulsars are born in supernova explosions, so it would be natural to look for
an explanation in the dynamics of the supernova~\cite{explosion}. However,
state-of-the-art 3-dimensional numerical calculations~\cite{fryer} show that
even the most extreme asymmetric explosions do not produce pulsar
velocities greater than 200~km/s.  Earlier 2-dimensional
calculations~\cite{sn_2D} claimed a maximal pulsar velocity up to 500~km/s
to be possible.  Of course, even this size of the kick was way too small to
explain the large population of pulsars with speeds above 1000~km/s.

Given the absence of a ``standard'' explanation, one is compelled to
consider alternatives, possibly involving new physics.  One of the reasons
why the standard explanation fails is because most of the energy is carried
away by neutrinos, which escape isotropically.  The remaining energy must
be distributed with a substantial asymmetry to account for the large pulsar
velocities.  In contrast, only a few per cent anisotropy in the
distribution of neutrinos, would give the pulsar a kick of required
magnitude.

Neutrinos are {\em produced } anisotropically, but they {\em escape} 
isotropically.  The asymmetry in production comes from the asymmetry in
the basic weak interactions in the presence of a strong magnetic field.
Indeed, if the electrons and other fermions are polarized by the magnetic
field, the cross section of the urca processes, such as $n+e^+
\rightleftharpoons p+ \bar \nu_e$ and $p+e^-\rightleftharpoons n+ \nu_e $,
depends on the orientation of the neutrino momentum.  Depending on the
fraction of the electrons in the lowest Landau level, this asymmetry can be
as large as 30\%, much more than one needs to explain the pulsar
kicks~\cite{drt}.  However, this asymmetry is completely washed out by
scattering of neutrinos on their way out of the star~\cite{eq}.

If, however, the same interactions produced a particle which had even
weaker interactions with nuclear matter than neutrinos, such a particle
could escape the star with an asymmetry equal its production asymmetry. 
It is intriguing that the same particle can the dark mater.  

The simplest realization of this scenario is a model that adds only one
singlet fermion to the Standard Model.  The SU(2)$\times$U(1) singlet, a
sterile neutrino, mixes with the usual neutrinos, for example, with the
electron neutrino.  Theoretical models can readily accommodate a sterile
neutrinos with desired properties~\cite{babu}.

For a sufficiently small mixing angle between $\nu_e$ and $\nu_s$, only one
of the two mass eigenstates, $\nu_1$, is trapped.  The orthogonal state,
%
$| \nu_2 \rangle = \cos \theta_m | \nu_s \rangle + \sin \theta_m | \nu_e
\rangle $, 
%
escapes from the star freely.  This state is produced in the same basic
urca reactions ($\nu_e+n\rightleftharpoons p+e^-$ and
$\bar\nu_e+p\rightleftharpoons n+e^+$) with the effective Lagrangian
coupling equal the weak coupling times $\sin \theta_m$.

We will consider two ranges of parameters, for which the $\nu_e \rightarrow
\nu_s$ oscillations occur on or off resonance. First, let us suppose that a 
resonant oscillation occurs somewhere in the core of the neutron star.
Then the asymmetry in the neutrino emission comes from shift in the
resonance point depending on the magnetic field~\cite{ks97}.  Second, we
will consider the off-resonance case, in which the asymmetry comes directly
from the weak processes, as described above~\cite{fkmp}.

Neutrino oscillations in a magnetized medium are described by an effective
potential which, in addition to the usual MSW potential has a
magnetic-field dependent term~\cite{magn}:   
\begin{equation}
V_{\rm eff} (B)  =  V_{B=0}
+\frac{e G_{_F}}{\sqrt{2}} \left ( \frac{3 N_e}{\pi^4} 
\right )^{1/3}
\frac{\vec{k} \cdot \vec{B}}{|\vec{k}|} \label{Vnumu}
\end{equation}
where $Y_e$ ($Y_{\nu_{\rm e}}$) is the ratio of the number density of electrons
(neutrinos) to that of neutrons, $\vec{B}$ is the magnetic field, 
$\vec{k}$ is the neutrino momentum, $V_0=10 \: \rm{eV} \: (\rho/10^{14} g
\, cm^{-3} )$.  The magnetic field dependent term in equation (\ref{Vnumu})
arises from polarization of electrons and {\em not} from a neutrino
magnetic moment, which is small and which we will neglect.

The condition for resonant MSW oscillation $\nu_i \leftrightarrow
\nu_s$ is
\begin{equation}
 V_{\rm eff} (B)= 
\frac{\Delta m^2}{2 k} \: \cos \, 2\theta . 
\label{res}
\end{equation}

In the presence of the magnetic field, the condition (\ref{res}) is
satisfied at different distances $r$ from the center, depending on the
value of the $(\vec{k} \cdot \vec{B})$ term in (\ref{Vnumu}). The average
momentum carried away by the neutrinos depends on the temperature of the
region from which they escape.  The deeper inside the star, the higher is
the temperature during the neutrino cooling phase.  Therefore, neutrinos
coming out in different directions carry momenta which depend on the
relative orientation of $\vec{k}$ and $\vec{B}$.  This causes the following
degree of asymmetry
in the momentum distribution~\cite{ks97,k_expl}: 
\begin{equation}
\frac{\Delta k}{k} = \frac{4 e\sqrt{2}}{\pi^2} \: 
\frac{\mu_e \mu_n^{1/2}}{m_n^{3/2}T^2} \ B =
0.01 \left ( \frac{B}{3\times 10^{15} {\rm G} }\right )
\label{dk2}
\end{equation}
if the neutrino oscillations take place in the core of the neutron star, at
density of order $10^{14} \, {\rm g\,cm^{-3}}$.  The neutrino oscillations
take place at such a high density if one of the neutrinos has mass in the
keV range, while the other one is much lighter.  The magnetic field of the
order of $10^{15}-10^{16}$~G is quite possible inside a neutron star, where
it is expected to be higher than on the surface.  (In fact, some neutron
stars, dubbed magnetars, appear to have surface magnetic fields of this
magnitude.)

Some comments are in order.  First, a similar kick mechanism, based
entirely on active neutrino oscillations (and no steriles) could also work
if the resonant oscillations took place between the electron and muon (tau) 
neutrinospheres~\cite{ks96}.  This, however, would require the mass
difference between two neutrinos to be of the order of 100~eV, which is
ruled out.  Second, the neutrino kick mechanism was criticized incorrectly 
by Janka and Raffelt~\cite{jr}.  It was subsequently shown by several
authors~\cite{ks98,k_expl,barkovich} that Janka and Raffelt made several
errors in their calculation, which is why their estimates differ from
eq.~(\ref{dk2}).

\begin{figure}
\centerline{\epsfxsize 4.1in \epsfbox{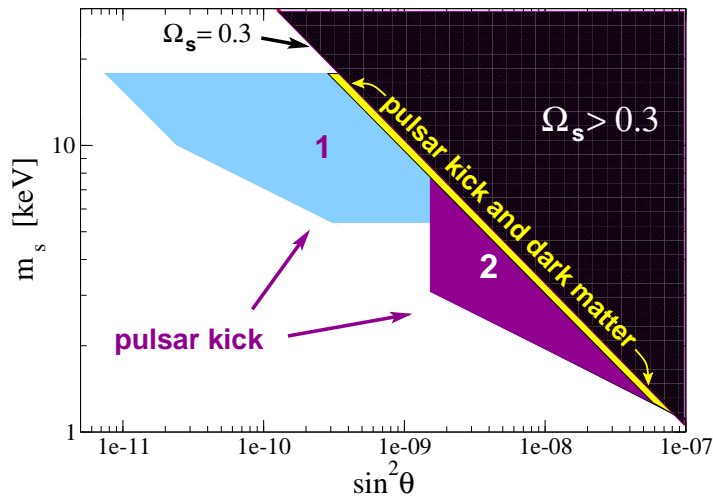}}   
\caption{The range of parameters for the sterile neutrino mass and mixing.
  Regions 1 and 2 correspond to parameters consistent with the pulsar kicks
  for (1) resonant and (2) off-resonant transitions, respectively.  Both
  regions overlap with a band in which the sterile neutrino is 
  dark matter.}
\end{figure}

For somewhat lighter masses, the resonant condition is not satisfied
anywhere inside the star.  In this case, however, the off-resonant
production of sterile neutrinos in the core can occur through ordinary urca
processes.  A weak-eigenstate neutrino has a $\sin^2\theta $ admixture
of a heavy mass eigenstate $\nu_2$.  Hence, these heavy neutrinos can be
produced in weak processes with a cross section suppressed by $\sin^2\theta
$. 

Of course, the mixing angle in matter $\theta_m$ is not the same as it is
in vacuum, and initially $\sin^2\theta_m \ll \sin^2\theta$.  However, as
Abazajian, Fuller, and Patel~\cite{Fuller} have pointed out, in the presence
of sterile neutrinos the mixing angle in matter quickly evolves toward its
vacuum value.  When $\sin^2\theta_m \approx \sin^2\theta$, the production 
of sterile neutrinos is no longer suppressed, and they can take a fraction
of energy out of a neutron star.  

Sterile neutrinos escape with a sizable asymmetry due to weak interactions
of fermions polarized by the magnetic field.  (Once again, we neglect a
neutrino magnetic moment and consider only the matter fermions.)  The
resulting asymmetry can explain the pulsar kicks if the mass and mixing
angle fall inside region 2 in Fig.~1.

The parameter space allowed for the pulsar kicks~\cite{fkmp} overlaps nicely
with that of dark-matter sterile neutrinos~\cite{Fuller,dw}.  Sterile
neutrinos in this range may soon be discovered~\cite{aft}.  Relic sterile
neutrinos with mass in the 1-20~keV range can decay into a lighter neutrino
and a photon.  The X-ray photons should be detectable by the X-ray
telescopes.  Chandra and XMM-Newton can exclude part of the parameter
space~\cite{aft}.  The future Constellation-X can probably explore the
entire allowed range of parameters.
In the event of a nearby supernova, the neutrino kick can produce gravity
waves that could be detected by LIGO and LISA~\cite{loveridge,cuesta}. 

Active-to-sterile neutrino oscillations can give a neutron star a kick.
However, if a black hole is born in a supernova, it would not receive a
kick, unless it starts out as a neutron star and becomes a black hole
later, because of accretion. (The latter may be what happened in SN1987A,
which produced a burst of neutrinos, but no radio pulsar.) If the central
engines of the gamma-ray bursts are compact stars, the kick mechanism
acting selectively on neutron stars and not black holes could probably
explain the short bursts as interrupted long bursts~\cite{k_semi}.

To summarize, the nature of cosmological dark matter is still unknown.  We
know that at least one particle beyond the Standard Model must exist to
account for dark matter.  This particle may come as part of a ``package'' if
supersymmetry is right.  However, it may be that the dark matter particle
is simply an SU(2)$\times$U(1) singlet fermion, which has a small mixing
with neutrinos.  In the latter case, the same dark-matter particle would be
emitted anisotropically from a supernova with an asymmetry sufficient to
explain the pulsar kick velocities.

This work was supported in part by the {DOE} grant {DE-FG03-91ER40662} and
the {NASA ATP} grant {NAG5-13399}.

\end{document}